\documentstyle[fleqn,twoside,espcrc2]{article}
\newcommand{\AmS}{{\protect\the\textfont2
  A\kern-.1667em\lower.5ex\hbox{M}\kern-.125emS}}

\def\PL{{\it Phys.Lett.} } 
 
\def\PRL{{\it Phys.Rev.Lett.} }

\def\PTP{{\it Progr.Theor.Phys.} }

\def\gappeq{\mathrel{\rlap {\raise.5ex\hbox{$>$}} {\lower.5ex\hbox{$\sim$}}}}

\def\lappeq{\mathrel{\rlap{\raise.5ex\hbox{$<$}} {\lower.5ex\hbox{$\sim$}}}}

\def\beq{\begin{equation}}
\def\eeq{\end{equation}}
\def\bea{\begin{eqnarray}}
\def\eea{\end{eqnarray}}

\title{
\vspace{-.5cm}
\begin{flushright}
{\normalsize CERN-TH-97-195 \\ \vspace{-.2cm}
hep-ph/9708437}
\end{flushright}
\vspace{.5cm}
HERA DATA AND LEPTOQUARKS IN SUPERSYMMETRY}
\author{G. Altarelli \address{Theoretical Physics Division,CERN,
CH-1211 Geneva 23, \\ and  Terza Universit\`a di Roma, Rome, Italy}}
       
\begin{document}

\begin{abstract}
I present a concise review of the possible evidence for new physics at HERA and of the recent work towards a
theoretical interpretation of the signal. It is not clear yet if the excess observed at large $Q^2$ is a
resonance or a continuum (this tells much about the quality of the signal). I discuss both possibilities. For the
continuum case one considers either modifications of the quark structure functions or contact terms. In the case of
a resonance, a leptoquark, the most attractive possibility that is being studied is in terms of s-quarks with
R-parity violation. In writing this script I updated the available information to include the new data and the
literature presented up to August 1, 1997.   
 
\end{abstract}
\maketitle
\section{Introduction}

The HERA experiments H1~\cite{H1} and ZEUS~\cite{ZEUS}, recently updated in ref.~\cite{LP}, have 
reported an excess of deep-inelastic $e^+p$
scattering events at large values of $Q^2\gappeq1.5 \times 10^4$ GeV$^2$, in a
domain not previously explored by other
experiments. The total $e^+p$ integrated luminosity was of 14.2 +9.5 = 23.7 pb$^{-1}$,
at H1 and of 
20.1+13.4 = 33.5 pb$^{-1}$ at ZEUS. The first figure refers to the data before the '97 run ~\cite{H1}~\cite{ZEUS},
while the second one refers to part of the continuing '97 run, whose results were presented at the LP'97 Symposium
in Hamburg at the end of July ~\cite{LP}. In the past, both experiments collected  about $1~pb^{-1}$ each with an
$e^-$ beam. A very schematic description of the situation is as follows. At
$Q^2\gappeq 1.5~10^4$ GeV$^2$ in the neutral current channel (NC), H1 observes 12+6 = 18 events while about 5+3 = 8 were
expected and ZEUS observes 12 + 6 = 18 events with about 9 + 6 =15 expected. In the charged current channel (CC),
in the same range of $Q^2$, H1 observes 4+2 = 6 events while about 1.8+1.2 = 3 were expected and ZEUS observes 3 + 2
= 5 events with about 1.2 + 0.8 = 2 expected. The distribution of the first H1 data suggested a resonance in the NC
channel. In the interval
$187.5<M<212.5~GeV$, which corresponds to $x\simeq 0.4$, and $y>0.4$, H1 in total finds 7 + 1 = 8 events with about
1 + 0.5 = 1.5 expected. But in correspondence of the H1 peak ZEUS observes a total of 3 events, just about the
 expected number. In the domain $x>0.55$ and $y>0.25$ ZEUS observes 3 + 2 events with about 1.2 + 0.8 = 2
expected. But in the same domain H1 observes only 1 event in total, more or less as expected. 

We see that with new statistics
the evidence for the signal remain meager. The perplexing features of the original data did not improve. First, there is a
problem of rates. With more integrated luminosity than for H1, ZEUS sees about the same number of events in
both the NC and CC channels. Second, H1 is suggestive of a resonance (although the evidence is now less than it was)
while ZEUS indicates a large $x$ continuum (here also the new data are not more encouraging).  The
difference could in part, but apparently not completely ~\cite{Dr-Ber},
be due to the different methods of mass reconstruction used by the two
experiments, or to fluctuations in the event characteristics. Of course,
at this stage, due to the limited
statistics, one cannot exclude the possibility that the whole effect is
a statistical fluctuation. All these issues will hopefully
be clarified by the continuation of data taking. Meanwhile, it is important to
explore possible
interpretations of the signal, in particular with the aim of identifying
additional
signatures that might eventually be able to discriminate between different
explanations of the reported excess.

\section{Structure Functions}

Since the observed excess is with respect to the Standard Model (SM)
expectation based on the QCD-improved parton model, the first question is
whether the effect could be explained by some
inadequacy of the conventional analysis without invoking new physics
beyond the SM. In the
somewhat analogous
case of the apparent excess of jet production at large
transverse energy $E_T$
recently observed by the CDF collaboration at the Tevatron~\cite{CDF}, it
has been argued~\cite{CTEQ} that a substantial
decrease in the discrepancy can be obtained by modifying the gluon parton
density at large values
of $x$ where it has not been measured directly. New results ~\cite{E756} on large $p_T$ photons appear to cast
doubts on this explanation because these data support the old gluon density and not the newly proposed one. In the
HERA case, a similar explanation appears impossible, at least for the H1 data. Here quark densities are involved and
they are well known at the same $x$ but smaller $Q^2$ ~\cite{tung},~\cite{rock}, and indeed the theory fits the data
well there. Since the QCD evolution is believed to be safe in the relevant region of $x$, the proposed strategy is
to have, at small $Q^2$, a new component in the quark densities at very large $x$, beyond the measured region,  which
is then driven at smaller $x$  by the evolution and contributes to HERA when $Q^2$ is sufficiently large ~\cite{tung}.
One possible candidate for a non perturbative effect at large $x$ is intrinsic charm ~\cite{charm}. However it
turns out that a large enough effect is only conceivable at very large $x$, $x\gappeq 0.75$, which is too large
even for ZEUS. The compatibility with the Tevatron is also an important constraint. This is because $ep$ scattering
is linear in the quark densities, while $p \bar p$ is quadratic, so that a factor of 1.5-2 at HERA implies a large
effect also at the Tevatron. In addition, many possibilities including
intrinsic charm (unless
$\bar c \not = c$ at the relevant $x$ values ~\cite{Thomas}) are excluded from the HERA data in the CC channel
~\cite{Babu2}. In conclusion, it is a fact that nobody sofar was able to even roughly fit the data. This
possibility is to be kept in mind if eventually the data will drift towards the SM and only a small
excess at particularly large $x$ and $Q^2$ is left with comparable effects in NC and CC, with $e^+$ or $e^-$
beams.

\section{Contact Terms}
Still considering the possibility that the observed excess is a
non-resonant continuum, a
rather general approach in terms of new physics is to interpret the HERA excess as due to an
effective four-fermion ${\bar e} e {\bar q} q$ contact
interaction~\cite{Eichten83}
with a scale $\Lambda$ of order a few TeV. It is
interesting that a similar contact term of the
${\bar q} q {\bar q} q$ type, with a scale of exactly the same order of
magnitude, could also reproduce
the CDF excess in jet production at large $E_T$ ~\cite{CDF}. (Note,
however, that this interpretation is not strengthened by
more recent data on
the dijet angular distribution~\cite{CDFangle}).
One has studied in detail ~\cite{ALTARELLI97}~\cite{CT} vector contact terms of the general form
\beq
\Delta L=\frac{4\pi\eta_{ij}}{(\Lambda^\eta_{ij})^2} \; \bar
e_i\gamma^{\mu}e_i \; \bar q_j\gamma_{\mu}q_j
\label{0}.
\eeq 
with $i,j=L,R$ and $\eta$ a $\pm$ sign. Strong limits on these
contact
terms
are provided by LEP2~\cite{LEP2} (LEP1 limits also have been considered but are less constraining
~\cite{LEP1,ELS}), Tevatron~\cite{CT-CDF} and atomic parity violation (APV) experiments~\cite{wood}. The constraints
are even more stringent for scalar or tensor contact terms. APV limits essentially exclude all relevant $A_eV_q$
component. The CDF limits on Drell-Yan production are particularly constraining. Data exist both for electron and
muon pairs up to pair masses of about 500 GeV and show a remarkable $e-\mu$ universality and agreement with the SM.
New LEP limits  (especially from LEP2) have been presented ~\cite{LEP2}. In general it would be possible to obtain
a reasonably good fit of the HERA data, consistent with the APV and the LEP limits, if one could skip the CDF limits
~\cite{barg}.  But, for example, a parity conserving combination
$({\bar e}_L\gamma^\mu e_L)(  {\bar u}_R\gamma_\mu u_R) +
( {\bar e}_R\gamma^\mu e_R )( {\bar u}_L\gamma_\mu u_L)$
with $\Lambda^+_{LR}=\Lambda^+_{RL}\sim 4$ TeV  still leads to a marginal fit to
the HERA
data and is compatible with all existing
limits~\cite{barg,Dib}. Because we expect contact terms to satisfy $SU(2)\bigotimes
U(1)$, as they reflect physics at large energy scales, the above phenomenological form is
to be modified into $\bar L_L \gamma_{\mu} L_L (\bar u_R \gamma^{\mu} u_R +\bar d_R \gamma^{\mu} d_R) + \bar e_R
\gamma_{\mu} e_R \bar Q_L \gamma^{\mu} Q_L)$, where L and Q are doublets ~\cite{car}. This form is both gauge
invariant and parity conserving. We took into account the requirement that contact terms corresponding to CC are
too constrained to appear. More general fits have also been performed ~\cite{barg}.

In conclusion, contact terms are severely constrained but not excluded. The problem of generating the
phenomenologically required contact terms from some form of new physics at larger energies is far from trivial
~\cite{car,stru}. Note also that contact terms require values of
$g^2/\Lambda^2 \sim 4\pi/(3-4~{\rm TeV})^2$, which would imply a very strong
nearby
interaction. Alternatively, for $g^2$ of the order of the $SU(3)\bigotimes SU(2)
\bigotimes U(1)$ couplings, $\Lambda$ would fall below 1~TeV, where the contact
term description is inadequate. We recall that the effects of contact terms should be present in both the
$e^+$ and the
$e^-$ cases with comparable intensity. Definitely contact terms cannot produce a CC signal~\cite{alta2}, as we shall
see, and no events with isolated muons and missing energy.

\section{Leptoquarks}
I now focus on the possibility of a resonance with $e^+ q$ quantum
numbers, namely a leptoquark ~\cite{ALTARELLI97,Buch86,lqlimits,HERAlq,Bluemlein96,mont,donc}, of mass $M\sim
190-210$~GeV, according to H1. The most obvious possibility is that the production at HERA occurs from valence $u$
or $d$ quarks, since otherwise the coupling would need to be
quite larger, and more difficult to reconcile with existing
limits. However production from the sea is also considered. Assuming an
$S$-wave state, one may have either a scalar or a vector leptoquark. I only consider
here the first option, because vector leptoquarks are more difficult to reconcile with their apparent absence at
the Tevatron. The coupling
$\lambda$ for a scalar
$\phi$ is defined by
$\lambda \phi {\bar e}_L
q_R$ or $\lambda \phi {\bar e}_R q_L$,
The corresponding width is given
by
$\Gamma=\lambda^2M_{\phi}/16\pi$, and the production cross section on a free quark
is given in lowest order by $ \sigma \; = \; \frac{\pi}{4 s} \, \lambda^2\;$ .

Including also the new '97 run results, the combined H1 and ZEUS data, interpreted in terms of scalar leptoquarks
lead to the following list of couplings ~\cite{ALTARELLI97,Mlm,Kunszt97}:
\bea
e^+u~~~~~~~~~~~~~~~\lambda \sqrt{B}\sim 0.017-0.025\nonumber\\
e^+d~~~~~~~~~~~~~~~\lambda \sqrt{B}\sim 0.025-0.033\nonumber\\
e^+s~~~~~~~~~~~~~~~~~~\lambda \sqrt{B}\sim 0.15-0.25\label{1}
\eea
where B is the branching ratio into the e-q mode. By s the strange sea is meant. For comparison note that
the electric charge is $e=\sqrt{4\pi \alpha}\sim0.3$. Production via
$e^+\bar u$ or
$e^+\bar d$ is excluded by the fact that in these cases the production in $e^-u$ or
$e^- d$ would be so copious that it should have shown up in the small luminosity already collected in the $e^-p$
mode. The estimate of $\lambda$ in the strange sea case is merely indicative due to the large uncertainties on the
value of the small sea densities at the relatively large values of $x$ relevant to the HERA data. The width is in all
cases narrow with respect to the resolution: for $B\sim 1/2$ we have
$\Gamma
\sim 4-16~$MeV for valence and $350-1000~$MeV for sea densities.

It is important to notice that
improved data from  CDF and D0~\cite{E756} on one side and from
APV~\cite{wood} and LEP ~\cite{LEP2}
on the other considerably reduce the window for leptoquarks. Consistency with the Tevatron, where
scalar leptoquarks are produced via model-independent (and $\lambda$-independent) QCD processes with potentially
large rates, demands a value of B sizeably smaller than 1. In fact, the most recent NLO estimates of the squark and
leptoquark production cross sections~\cite{Spira96,Spira97} allow to estimate that at
200~GeV approximately 6--7 events with $e^+e^-jj$ final states
should be present in the combined CDF and D0 data
sets. For $B = 1$ the CDF limit
is 210~GeV, the latest D0 limit is 225~GeV at 95\%CL.
The combined CDF+D0 limit
is 240~GeV at 95\%CL ~\cite{E756}. 
We see that for consistency one should impose:
\beq
B \lappeq 0.5-0.7\\ \label{6}
\eeq
Finally, the case of a 200~GeV vector leptoquark 
is most likely totally ruled out by the
Tevatron data, since the production rate can be as much as a factor
of 10 larger than that of
scalar leptoquarks.

There are also lower limits on B, different for
production off valence or sea quarks, so that only a definite window for B is left in all cases. For production
off valence the best limit arises from APV ~\cite{wood}, while for the sea case it is obtained from recent LEP2 data
~\cite{LEP2}.

One obtains a limit from APV because the s-channel exchange amplitude for a leptoquark is equivalent at low
energies to an $(\bar e q)(\bar q e)$ contact term with amplitude proportional to $\lambda^2/M^2$. After Fierz
rearrangement a component on the relevant APV amplitude $A_eV_q$ is generated, hence the limit on $\lambda$. The
results are ~\cite{alta2}
\bea
e^+u~~~~~~~~~~~~~~~\lambda\lappeq 0.058\nonumber\\
e^+d~~~~~~~~~~~~~~~\lambda\lappeq 0.055\label{3}
\eea
The above limits are for $M=200$~GeV (they scale in proportion to M) and are obtained from the quoted error on
the new APV measurement on Cs. This error being mainly theoretical, one could perhaps take a more conservative
attitude and somewhat relax the limit. Comparing with the values for
$\lambda
\sqrt{B}$ indicated by HERA, given in eq.(\ref{1}), one obtains lower limits on B:
\bea
e^+u~~~~~~~~~~~~~~~B\gappeq 0.1-0.2\nonumber\\
e^+d~~~~~~~~~~~~~~~B\gappeq 0.2-0.4
\label{33}
\eea
For production off the strange sea quark the upper limit on $\lambda$ is obtained from LEP2 ~\cite{LEP2}, in that
the t-channel exchange of the leptoquark contributes to the process $e^+e^-\rightarrow s \bar s$ (similar limits
for valence quarks are not sufficiently constraining, because the values of $\lambda$ required by HERA are
considerably smaller). Recently new results have been presented by ALEPH, DELPHI and OPAL ~\cite{LEP2}. The best
limit is from ALEPH:
\beq
e^+s~~~~~~~~~~~~~~~~~~\lambda\lappeq 0.6\\
\label{4}  
\eeq
(OPAL finds $\lambda\lappeq 0.7$, DELPHI $\lambda\lappeq 0.9$). This, given eq.(\ref{1}), corresponds to  
\beq
e^+s~~~~~~~~~~~~~~~B\gappeq 0.05-0.2\\
\label{5}
\eeq
Recalling the Tevatron upper limits on B, given in eq.(\ref{6}),  we see from eqs. (\ref{33}) and (\ref{5}) that only a definite window for B is left
in all cases. 

Note that one given leptoquark cannot be present both in $e^+p$ and in $e^-p$ (unless it is produced from strange
quarks).

\section{S-quarks with R-parity Violation}

I now consider specifically leptoquarks and SUSY ~\cite{ALTARELLI97,ELS,Moha,Hewett,Kon,Dreiner,CR,HERARPV}. In general, in
SUSY one could consider leptoquark models without R-parity violation. It is sufficient to introduce
together with scalar leptoquarks also the associated spin-1/2 leptoquarkinos ~\cite{Moha}. In this way one has not
to give up the possibility that neutralinos provide the necessary cold dark matter in the universe. We find it more
attractive to embed a hypothetical leptoquark in the minimal supersymmetric extension of the 
SM~\cite{MSSM} with violation of
$R$ parity~\cite{RPth}. The connection with the HERA events
has been more recently invoked in ref.~\cite{ALTARELLI97,Kon,CR}. The
corresponding superpotential can be written in the form
\bea
W_R &\equiv& \mu_i H L_i + \lambda_{ijk} L_i L_j E^c_k
+ \lambda'_{ijk} L_i Q_j D^c_k + \nonumber \\
&& \lambda ''_{ijk} U^c_i D^c_j D^c_k ~,
\label{WR}
\eea
where $H, L_i, E^c_j, Q_k, (U,D)^c_l$ denote superfields for the
$Y= 1/2$ Higgs doublet, left-handed lepton doublets,
lepton singlets, left-handed quark doublets and quark
singlets, respectively. The indices $i,j,k$ label the three generations
of quarks and leptons. 
Furthermore, we assume the absence of the $\lambda''$ couplings, so as to
avoid rapid baryon decay, and the $\lambda$ couplings play no r\^ole
in our analysis.

The squark production mechanisms permitted by the $\lambda'$
couplings in (\ref{WR}) include $e^+ d$ collisions to form
${\tilde u}_L, {\tilde c}_L$ or $\tilde t_L$, which
involve valence $d$ quarks, and various collisions of the
types $e^+ {d_i}$ ($i=2,3$) or $e^+ {\bar u_i}$ ($i=1,2,3$) which involve sea quarks. A careful analysis \cite{ALTARELLI97} leads to
the result that the only processes that survive after taking into account existing low energy limits are
\bea
e^+_Rd_R \rightarrow \tilde c_L \nonumber\\
e^+_Rd_R \rightarrow \tilde t_L \nonumber\\
e^+_Rs_R \rightarrow \tilde t_L 
\label{7}
\eea
For example $e^+_Rd_R \rightarrow \tilde u_L$ is forbidden by data on neutrinoless double beta decay which imply 
\cite{Hirsch}
\beq
\vert \lambda'_{111} \vert < 7 \times 10^{-3} \left( {m_{\tilde q} \over 200
~\hbox{GeV}}\right)^2
\left( {m_{\tilde g} \over 1 ~\hbox{TeV}}\right)^{1 \over 2}~.
\label{betabeta}
\eeq
where $m_{\tilde q}$ is the mass of the lighter of
${\tilde u}_L$ and ${\tilde d}_R$, and $m_{\tilde g}$ is the gluino mass.

It is interesting to note ~\cite{Kon} that the left s-top could be a superposition of two mass eigenstates
$\tilde t_1,\tilde t_2$, with a difference of mass that can be large as it is proportional to $m_t$:
\beq
\tilde t_L = \cos\theta_t~\tilde t_1~+~\sin\theta_t~\tilde t_2\\
\label{8}
\eeq
where $\theta_t$ is the mixing angle. With $m_1 \sim 200$~GeV, $m_2 \sim 230$~GeV and $\sin^2\theta_t \sim 2/3$
one can obtain a broad mass distribution, more similar to the combined H1 and ZEUS data. (But with the present data
one has to swallow that H1 only observes $\tilde t_1$ while ZEUS only sees $\tilde t_2$!). However, the presence
of two light leptoquarks makes the APV limit more stringent. In fact it becomes
\beq
B>B_{\infty} [1 + \tan^2{\theta_t}\frac{m^2_1}{m^2_2}]\\
\label{99}
\eeq
Thus, for the above mass and mixing choices, the above quoted APV limit
[$B_\infty$ is given in eq. (\ref{33})]  must be relaxed invoking a
larger theoretical uncertainty on the Cs measurement.

Let us now discuss ~\cite{ALTARELLI97} if it is reasonable to expect that $\tilde
c$ and
$\tilde t$ decay satisfy the bounds on the branching ratio B. A virtue of s-quarks as leptoquark is that
competition of R-violating and normal decays ensures that in general $B<1$.

In the
case of ${\tilde c}_L$, the
most important possible decay modes are
the $R$-conserving channels ${\tilde c}_L\rightarrow c \chi^0_i$ ($i=1,..,4$) and
${\tilde c}_L\rightarrow s \chi^+_j$ ($j=1,2$), and
the $R$-violating channel ${\tilde c}_L\rightarrow d e^+$, where
$\chi^0_i, \chi^+_j$ denote neutralinos and charginos, respectively. In this case it has been shown \cite{ALTARELLI97} that, if one
assumes that
$m_{\chi^+_j} > 200$ GeV, then , in a
sizeable domain of the parameter space, the neutralino mode can be sufficiently suppressed so that
$B\sim 1/2$ as required (for example, the couplings of a higgsino-like neutralino are suppressed by the small
charm mass). 

In the case of ${\tilde t}_L$,
it is interesting to notice that the
neutralino decay mode ${\tilde t}_L \to t \chi^0_i$ is kinematically closed
in a natural way. Thus, in order to obtain a large value of B in the case of s-top production off d-quarks, in spite of
the small value of $\lambda$, it is sufficient to require that all charginos are heavy enough to forbid the
decay ${\tilde t}_L \to b \chi^+_j$. However, we do not really want to obtain B too close to 1, so that in this
case some amount of fine tuning is required. Or, with charginos heavy, one could invoke other decay channels as,
for example, $\tilde t \rightarrow \tilde b W^+$ ~\cite{kon2}. But the large splitting needed between $\tilde t$
and 
$\tilde b$ implies problems with the $\rho$-parameter of electroweak precision tests, unless large mixings in both
the s-top and s-bottom sectors are involved and their values suitably chosen. 
In the case of s-top production off s-quarks, values around $B\sim 1/2$ are rather natural because of
the larger value of $\lambda$, which is of the order of the gauge couplings \cite{ALTARELLI97,ELS}.

The interpretation of HERA events in terms of s-quarks with R-parity violation requires a very peculiar
family and flavour structure~\cite{Barbieri97}. The flavour problem is that there are very strong limits on
products of couplings from absence of FCNC. The unification problem is that nucleon stability poses even stronger
limits on products of
$\lambda$ couplings that differ by the exchange of quarks and leptons which are treated on the same footing in
GUTS. However it was found that the unification problem can be solved and the required
pattern can be embedded in a grand unification framework~\cite{Barbieri97}. The already intricated problem
of the mysterious texture of masses and couplings is however terribly enhanced in these scenarios.

\section{Charged Current Events}

In the Introduction, I have mentioned that in the CC channel at $Q^2\gappeq 1.5~10^4$ GeV$^2$ H1 and ZEUS see a total of 11 events with 5
expected. The statistics is even more limited than in the NC case, so one cannot at the moment derive any firm
conclusion on the existence and on the nature of an excess in that channel. However, the presence or absence of a
simultaneous CC signal is extremely significant for the identification of the underlying  physical effect (as it would
also be the case for
the result of a comparable run with an $e^-$ beam, which however is further
away in time). In view of this, 
I now briefly discuss the implications for the CC channel of the various
proposed solutions of the HERA effect~\cite{alta2,kon2,babu3,care}. It is found that in most of the
cases the CC signal is not expected to arise. But if it is present at a
comparable rate as for the NC signal, the corresponding indications are very
selective.

To see that contact terms cannot work recall that for them it is natural to assume the validity of the
$SU(2)\bigotimes U(1)$ symmetry, because they are associated with physics
at a
large energy scale. In the
$SU(2)\bigotimes U(1)$ limit, restricting us to family diagonal quark
currents in order to minimise problems with the occurrence of flavour
changing neutral currents, the only possible vector contact term with
valence quarks (and no Cabibbo suppression) is of the form
\beq
\Delta L_{CC}=\frac{4\pi\eta}{\Lambda^2_{\eta}}
{\bar e}_L\gamma^\mu \nu_L {\bar u}_L\gamma_\mu d^\prime_L + {\rm h.c.}
\label{due}
\eeq {\it i.e.} the product of two isovector currents.
Here $d_L^\prime$ is the left-handed d-quark current eigenstate, related
to the mass eigenstate by the Cabibbo-Kobayashi-Maskawa (CKM) matrix. It is simple to see that such
terms cannot have a sufficent magnitude. In fact the
scale
$\Lambda$ associated with this operator is too strongly constrained to produce any measurable effect at HERA. The
constraints arise from at least two
experimental facts: lepton-hadron universality of weak charged currents
and electron-muon universality in charged-pion decays. The corresponding lower limits on $\Lambda$ exceed
$10$~TeV in all cases ~\cite{alta2}. 

The possible scalar or tensor currents arising from an $SU(2)\otimes U(1)$
invariant theory which can contribute to valence-parton CC processes are
\bea
{\cal L}&=&
\frac{4 \pi}{\Lambda_S^2}({\bar e}_R \nu_L)( {\bar u}_R d_L)+
\frac{4 \pi}{\Lambda_{S^\prime}^2}({\bar e}_R \nu_L)( {\bar u}_L d_R)+
\nonumber \\
&&\frac{4 \pi}{\Lambda_T^2}({\bar e}_R\sigma^{\mu \nu} \nu_L)( {\bar u}_R
\sigma_{\mu \nu}d_L) ~,
\eea
while the operator $({\bar e}_R\sigma^{\mu \nu} \nu_L)( {\bar u}_L
\sigma_{\mu \nu}d_R)$ identically vanishes. The scalar interactions are
strongly limited by $e$--$\mu$ universality in pion decays~\cite{sha},
because
they do not lead to electron-helicity suppression, in contrast with
the SM case. The lower limit om $\Lambda$ is about $500$~TeV~\cite{sha}. The tensor interaction can be dressed into
a scalar interaction of effective strength~\cite{vol}
\beq
\frac{1}{\Lambda_{S~eff}^2}\simeq
-\frac{\alpha}{\pi}\log\left( \frac{\Lambda_T^2}{M_W^2}\right) ~
\frac{1}{\Lambda_T^2}~,
\eeq
with the
exchange of a photon between the electron and the quark fields. The upper limit remains sufficiently strong to
prevent these terms from contributing as well.

Considering now also  CC processes involving sea quarks ~\cite{alta2}, we can
introduce a contact term for second generation quarks
\beq
\Delta L_{CC}=\frac{4\pi\eta}{\Lambda^{(2)2}_{\eta}}(
{\bar e}_L\gamma^\mu \nu_L)( {\bar c}_L\gamma_\mu s^\prime_L) + {\rm h.c.}
\label{charm}
\eeq
Clearly since the strange sea in the proton is small one needs relatively
small values of $\Lambda$ in order to produce a sufficiently large effect.
A detailed study shows that one needs $\Lambda\sim0.8-1$~TeV with
$\eta=-1$ in order to obtain an increase by a factor of two with respect
to the SM at $Q^2 \gappeq 15000$~GeV$^2$. But bounds on the scales $\Lambda_{\eta}^{(2)}$ derived
from
lepton universality in $D$ decays~\cite{alta2,pdg} and, independently, from the unitarity of the CKM matrix, forbid such
low values.

In conclusion it appears very difficult
to accommodate a CC signal at HERA in the framework of contact terms.

Let us now consider a scalar leptoquark resonance that is coupled both to
$e^+d$ and to $\bar \nu u$ so that it can generate both NC and CC events from
valence (note that $e^+u$ has charge +5/3 and cannot go into $\bar \nu q$). Assuming that the symmetry under
$SU(2)\bigotimes U(1)$ is conserved, the virtual leptoquark exchange
gives a CC contribution to the low-energy effective Lagrangian of the form
\beq
{\cal L}=\frac{\lambda_u \lambda_d}{M^2}(\bar e_R d_L)(\bar u_R \nu_L)
+{\it h.c.}
\label{9}
\eeq
Here $\lambda_u$ and $\lambda_d$ are the (real) couplings of a leptoquark with
mass $M$ to the $\bar \nu_L u_R $ and $\bar e_R d_L$ currents,
respectively.
This interaction corresponds to the transition $e^+_L d_L \rightarrow \bar
\nu_R
u_R$
which has $T=-1/2$ both in the initial and final states.
At low energies, the leptoquark exchange induces a contribution to
$\pi \to e \bar \nu$ which is not helicity suppressed. It is simple to verify that
this clearly excludes any observable CC signal.
An alternative is to break $SU(2)\bigotimes U(1)$ and assume that the
leptoquark exchange induces an effective interaction of the form
\beq
{\cal L}=\frac{\lambda_u \lambda_d}{M^2}(\bar e_L d_R)(\bar u_R \nu_L)
+{\it h.c.}
\label{10}
\eeq
Note that in the transition $e^+_R d_R \rightarrow \bar \nu_R u_R$ the
initial
state has $T=+1/2$ while the final state has $T=-1/2$. In this case the low
energy
effective interaction gives a contribution to $\pi \rightarrow e \bar \nu$
which
is helicity suppressed and so can be acceptable, as it can be checked.
Since $SU(2)\otimes U(1)$ is broken only by the Higgs vacuum
expectation value (VEV), the leptoquark couplings could violate
gauge invariance if the leptoquark couples to the
quark-lepton current through some higher-dimensional operator ~\cite{alta2} and/or if the breaking induces a mixing
between two leptoquarks of different electroweak properties ~\cite{babu3}.

Another viable alternative is a leptoquark which couples simultaneously to
the $\bar e_R^{(1)} q_L^{(1)}$ and $\bar \ell_L^{(i)} u_R^{(2)} $ currents
($i=1,2,3$).
Here we have specified the generation indices of the different fields.
If CC events were observed at HERA and such a leptoquark was responsible
for them, we expect the striking signature of leptonic $D$ decays with
rates much larger than in the SM. In the case of a leptoquark produced
in the $e^+ s$ channel,
the possibility of a ${\bar \nu} c$ final state is still allowed.
This leads to a remarkable signature in leptonic $D_s$ decays
\bea
BR(D_s^-\to e^-\bar \nu)&\simeq& 6\times 10^{-3}\frac{{\cal B}_{\nu
c}}{(1-{\cal B}_{\nu c})^3}\nonumber \\
&&\left( \frac{200~{\rm GeV}}{M}\right)^4
~i=1,2~.\nonumber\\
\eea
We are not aware of any existing experimental limit on this quantity.

To conclude, we recall that
a leptoquark with branching ratio equal to 1 in
$e^+q$ is excluded by the recent Tevatron limits.
Therefore on one hand some branching fraction in the CC channel
is needed. On the other hand, we find that there is limited space
for the possibility that a leptoquark can generate a CC signal at HERA with one
single parton quark in the final state. This occurrence would indicate
$SU(2)\bigotimes U(1)$ violating couplings or couplings to a current containing the
charm quark.

A few mechanisms for producing CC final states from $\tilde c$ or $\tilde t$ have been proposed
~\cite{alta2,kon2,care}. In all cases $\tilde c$ or $\tilde t$ lead to multiparton final states. Since apparently
the CC candidates are all with one single jet, some strict requirements on the masses of the
participating
particles must be imposed so that some partons are too soft to be visible while others
coalesce into a single
visible jet. Consider for example the chain ~\cite{alta2}
\beq
\tilde c \rightarrow c \chi^0\rightarrow c \nu \tilde{\nu}\rightarrow c \nu d \bar s\\ \label{que}
\eeq
where in the last step the R-violating coupling $\tilde{\nu} \rightarrow d \bar s$ is involved which, by gauge
symmetry and supersymmetry, has the same coupling $\lambda$ as the $ed \rightarrow \tilde c$ coupling. In order
for the c quark to be invisible the neutralino mass must be sufficiently close to the $\tilde c$ mass. For the d
and $\bar s$ partons to coalesce in a single jet, the s-neutrino mass must be small, close to the LEP2 limit
(actually if the ALEPH 4-jet events were true, they could be a manifestation of light s-neutrinos or s-leptons). A
similar chain could also lead to charged leptons plus missing energy in the final state. 

In a different mechanism
without light s-leptons one can use s-bottom decays, like in the chain ~\cite{care}.
\beq
\tilde t \rightarrow b \chi^+\rightarrow b c \tilde{\bar b}\rightarrow b c \nu \bar d \\ \label{qui}
\eeq
Here too the coupling $\tilde b \rightarrow d \bar{\nu}$ is implied by the $ed \rightarrow \tilde t$ coupling. 
In this case, in
order not to observe the b and c quark jets in the final state, one needs that the masses of $\tilde t$, charginos
and s-bottom are close and in decreasing order.

In conclusion, s-quarks with R-parity violating decays could produce CC events or events with charged leptons and
missing energy. The observation of such events would make the model much more constrained. 

It is a pleasure for me to thank Professors Mirjam Cvetic and Paul Langacker for
their kind invitation and pleasant hospitality.

\def\ijmp#1#2#3{{\it Int. Jour. Mod. Phys. }{\bf #1~}(19#2)~#3}
\def\pl#1#2#3{{\it Phys. Lett. }{\bf B#1~}(19#2)~#3}
\def\zp#1#2#3{{\it Z. Phys. }{\bf C#1~}(19#2)~#3}
\def\prl#1#2#3{{\it Phys. Rev. Lett. }{\bf #1~}(19#2)~#3}
\def\rmp#1#2#3{{\it Rev. Mod. Phys. }{\bf #1~}(19#2)~#3}
\def\prep#1#2#3{{\it Phys. Rep. }{\bf #1~}(19#2)~#3}
\def\pr#1#2#3{{\it Phys. Rev. }{\bf D#1~}(19#2)~#3}
\def\np#1#2#3{{\it Nucl. Phys. }{\bf B#1~}(19#2)~#3}
\def\mpl#1#2#3{{\it Mod. Phys. Lett. }{\bf #1~}(19#2)~#3}
\def\arnps#1#2#3{{\it Annu. Rev. Nucl. Part. Sci. }{\bf #1~}(19#2)~#3}
\def\sjnp#1#2#3{{\it Sov. J. Nucl. Phys. }{\bf #1~}(19#2)~#3}
\def\jetp#1#2#3{{\it JETP Lett. }{\bf #1~}(19#2)~#3}
\def\app#1#2#3{{\it Acta Phys. Polon. }{\bf #1~}(19#2)~#3}
 \def\rnc#1#2#3{{\it Riv. Nuovo Cim. }{\bf #1~}(19#2)~#3}
\def\ap#1#2#3{{\it Ann. Phys. }{\bf #1~}(19#2)~#3}
\def\ptp#1#2#3{{\it Prog. Theor. Phys. }{\bf #1~}(19#2)~#3}
\def\ZPC#1#2#3{{\sl Z.~Phys.} {\bf C#1}~(#3) #2}
\def\PTP#1#2#3{{\sl Prog. Theor. Phys.} {\bf #1}~(#3) #2}
\def\PRL#1#2#3{{\sl Phys. Rev. Lett.} {\bf #1}~(#3) #2}
\def\PRD#1#2#3{{\sl Phys. Rev.} {\bf D#1}~(#3) #2}
\def\PLB#1#2#3{{\sl Phys. Lett.} {\bf B#1}~(#3) #2}
\def\PREP#1#2#3{{\sl Phys. Rep.} {\bf #1}~(#3) #2}
\def\NPB#1#2#3{{\sl Nucl. Phys.} {\bf B#1}~(#3) #2}

\end{document}